# Fusion or Confusion? Potential and Challenges in Fusion of Onboard Sensors and V2X Data in Cooperative Perception

Amir Mohammadisarab[1], Miguel Sepulcre[1], Luca Lusvarghi[1], Sergei S. Avedisov[2],
Mohammad Irfan Khan[2], Takayuki Shimizu[2], Onur Altintas[2], Javier Gozalvez[1]
[1]Universidad Miguel Hernández de Elche (UMH), Spain, {amohammadisarab, msepulcre, llusvarghi, j.gozalvez}@umh.es
[2]Toyota Motor North America R&D, InfoTech Labs, Mountain View, California, USA,
{sergei.avedisov, mohammad.irfan.khan, takayuki.shimizu, onur.altintas}@toyota.com

***Abstract*—Connected Automated Vehicles (CAVs) utilize their onboard sensors to perceive the environment. The perception range and accuracy can be affected by adverse weather or non-line-of-sight conditions. Cooperative perception or sensor sharing can overcome these limitations by enabling CAVs to exchange sensor data, thus collectively enhancing their perception capabilities. Previous studies have shown the potential of cooperative perception, but limited attention has been given to the fusion of V2X data received through cooperative perception messages with onboard sensor information. The fusion process can be influenced by the quantity and quality of the V2X data. An increased volume of V2X data can reduce uncertainty in the perceived environment; however, when the data is noisy, it may compromise the accuracy of the fusion results. This study investigates the fusion of onboard sensor and V2X data in cooperative perception, and demonstrates that while perception can significantly improve as the V2X penetration rate increases, it can introduce a significant number of false positives if V2X data is not highly accurate. False positives result in the detection of ghost objects that do not actually exist. These ghost objects can, in turn, compromise safety and driving efficiency. Our analysis found that false positives or ghost objects can appear even with accurate V2X data. These findings highlight the challenges in cooperative perception and the importance of developing robust data fusion methods to enhance the reliability of cooperative perception. This is particularly relevant in light of ongoing standardization efforts, such as ETSI TS 103 324 on collective perception.**

***Keywords— V2X, cooperative perception, collective perception, sensor sharing, fusion, association, penetration rate, advanced driver assistance systems, ADAS, V2V.***

## I. Introduction

Cooperative (or collective) perception or sensor data sharing is a key enabler to enhance the perception of Connected and Automated Vehicles (CAVs) as well as human-driven vehicles using Vehicle-to-Everything (V2X) communication. Cooperative perception enables the exchange of information acquired by onboard sensors, which helps mitigate sensor limitations in terms of occlusion and field of view [1]. The standardization of cooperative perception by ETSI and SAE enables CAVs to exchange object- or track-level data (e.g., position, speed, and size) via V2X in cooperative perception messages. The received V2X data must be fused with onboard sensor data to enhance the perception capabilities of the CAVs. This type of fusion is known as high-level fusion, late fusion or track-to-track fusion, and considers that each sensor independently carries out filtering and tracking, and their outputs are then fused. Given the differences among vehicle positioning systems and the inherent delays associated with V2X communications, incoming object data must initially undergo spatial and temporal alignment to match onboard sensor information. Subsequently, association algorithms identify which objects, detected independently by onboard sensors and external CAVs, represent the same entity in the real world. Ultimately, a fusion process combines these matched observations into a single, coherent environmental representation [2].

The high-level fusion of data from multiple sensors requires adequate association and fusion algorithms, as demonstrated for the fusion of onboard sensors data of autonomous vehicles [3]. However, limited studies have been conducted on the fusion of onboard sensor data and V2X data. Certain studies, such as [4], have addressed the fusion of onboard sensor data with V2X data obtained through BSMs (Basic Safety Messages), which provide location information about the transmitter, but do not include information about the detected objects. Existing studies on the fusion of onboard sensors and V2X data for cooperative perception propose methods for spatial and temporal alignment. One example is the study in [5] that predicts the current object states using a constant turn rate and acceleration motion model and an unscented Kalman filter. Similar studies investigate the fusion performance under V2X communication latency, such as [6] that demonstrates that the increased latency does not significantly affect the error of the fusion process. Related studies on cooperative perception focused on ensuring scalability, for example, by efficiently selecting the objects to include in each cooperative perception message [7][8][9]. However, they often consider an ideal fusion process and thus could not study the impact of the quantity and quality of the V2X data on the performance of cooperative perception. An increased volume of V2X data can reduce uncertainty about the perceived objects, but when the data is noisy, it may degrade the accuracy of the fusion output.

In this context, this paper investigates the impact that the quantity and quality of V2X information has on the fusion process of onboard sensors data with V2X data for cooperative perception. Specifically, it investigates how an increased volume of V2X data—resulting from higher V2X penetration rates— impacts the perception when the information may be affected by sensing and positioning inaccuracies. To this end, the study analyzes varying levels of V2X data inaccuracies stemming from sensor noise – which affects the estimated

This work has been partly funded by MCIN/AEI/10.13039/501100011033 and the "European Union NextGenerationEU/PRTR" (TED2021-130436B-I00).

position and speed of detected objects- and from GNSS errors, which impact the reported position of the transmitters sending cooperative perception messages. Our study demonstrates the potential of cooperative perception to improve the perception range and the detection of real objects (true positives), but also the risk that V2X data generates false positives or ghost objects that do not exist, even when with accurate V2X data. These insights underscore the critical role of a robust and well-designed fusion process in maximizing the benefits of cooperative perception under real-world uncertainty.

## II. Fusion of Onboard Sensors and V2X Data

In this study, we implement a high-level fusion architecture designed to integrate onboard sensor data with information received through V2X communications. This approach builds upon and extends the architecture previously proposed in [10], adapting it to support cooperative perception. The implemented architecture, illustrated in Fig. 1, differentiates between two core stages of processing: sensor-level and fusion-level. Cooperative perception messages contain object-level information and are introduced at the sensor-level processing, where they are interpreted as virtual sensor inputs.

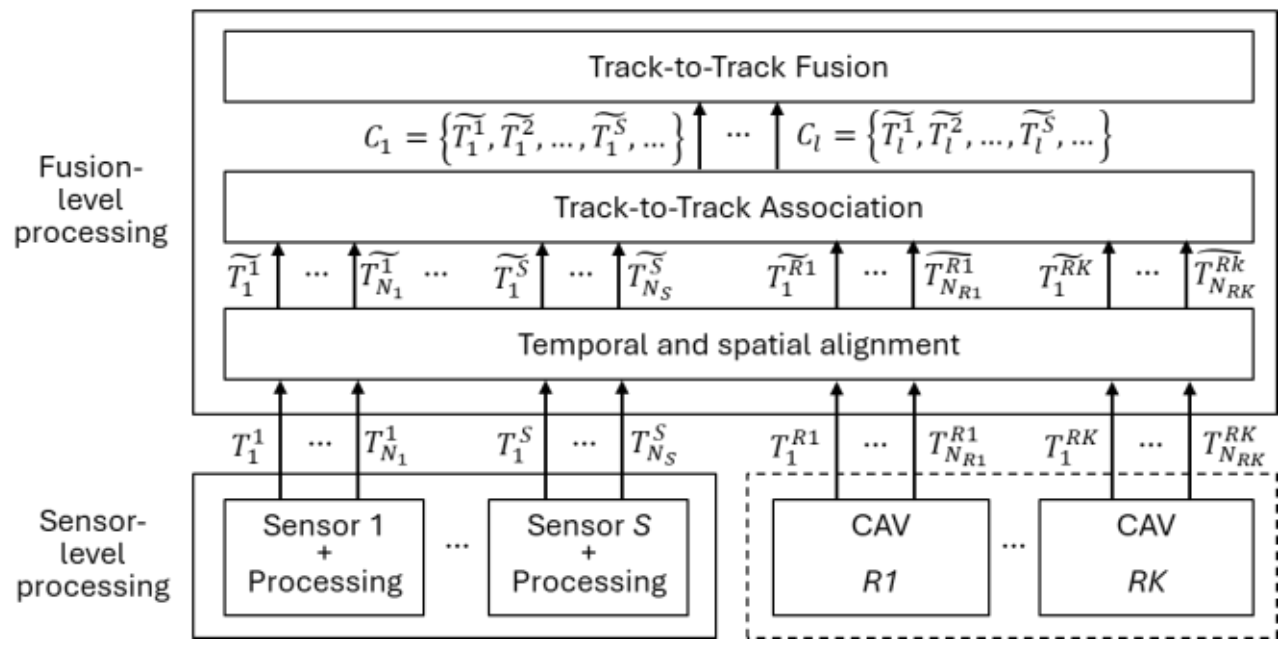

Fig. 1. High-level onboard sensors and V2X data fusion architecture.

At the sensor-level processing stage, each sensor independently observes its surroundings and carries out fundamental perception tasks including feature extraction, object classification, and object tracking. These sensors may be physical (e.g., camera, radar, LiDAR) or virtual (i.e., cooperative perception data received via V2X). In the proposed framework, an object $I$ detected by sensor $j$ is represented by a corresponding track $T_i^j$, which comprises a state vector $X_i^j$ and an associated covariance matrix $P_i^j$. Since all the processing is assumed to occur at time step $t$, for simplicity, the notations in this section omit the time index $t$. The state vector includes spatial and kinematic parameters, such as the object's position and velocity, while the covariance matrix—computed using a Kalman filter in this implementation—quantifies the uncertainty in the state estimate [11]. The output of each sensor is a list of $N_j$ tracks, all of which are fed into the subsequent fusion-level processing stage. The same format is used for tracks generated by onboard sensors and those received via V2X cooperative perception messages. An illustration of onboard sensor tracks at an ego vehicle, and V2X tracks received in cooperative perception messages is provided in Fig. 2.

Fusion-level processing begins with the task of temporal and spatial alignment. This step ensures that all incoming tracks are expressed within a common spatiotemporal reference, thereby enabling meaningful comparison and integration. Spatial alignment addresses discrepancies in sensor placement and orientation: onboard sensors may be mounted at different locations on the vehicle or aligned at various angles, and cooperative messages typically encode object positions relative to the sender's GNSS coordinates. Therefore, transforming these tracks—denoted as $\widetilde{T_i^j}$ in Fig. 1—into the ego vehicle's coordinate system must account for potential GNSS errors and sensor model differences. Temporal alignment compensates for asynchrony in track generation and reception, especially when messages are subject to network delay. In this work, we assume synchronized sensor data generation at both the ego vehicle and transmitting CAVs, justified by the relatively minor impact of latency once proper temporal alignment has been applied, as demonstrated in [6].

The next step in fusion-level processing is track-to-track association, which clusters tracks corresponding to the same real-world object (illustrated in Fig. 2). Each cluster, $C_i$, is assigned a unique identifier, $i$, ideally representing a single object. The association problem is typically formulated as an assignment optimization and commonly solved using methods such as Hungarian, Auction, or Greedy algorithms, depending on computational efficiency and accuracy requirements [12]. The Hungarian algorithm efficiently solves the assignment problem by constructing and iteratively improving a cost matrix to find the optimal pairing between observations and targets. Auction algorithms, on the other hand, use an iterative bidding process in which each observation "bids" for a target based on a cost criterion, converging to an optimal or near-optimal assignment. Greedy strategies construct the assignment step-by-step, always selecting the best available pairing at each iteration without considering the global optimum. This makes them computationally efficient but occasionally suboptimal. In this paper, we adopt the Greedy method detailed and extended in [13]. The Greedy approach calculates Euclidean distances between every pair of tracks and sequentially groups them into clusters by choosing pairs with the smallest distances, ensuring that tracks originating from the same sensor are not placed in the same cluster. In the extended Greedy approach proposed in [13], a merging phase is specifically added to combine tracks from different sensors, thus enhancing performance when handling data from multiple sources.

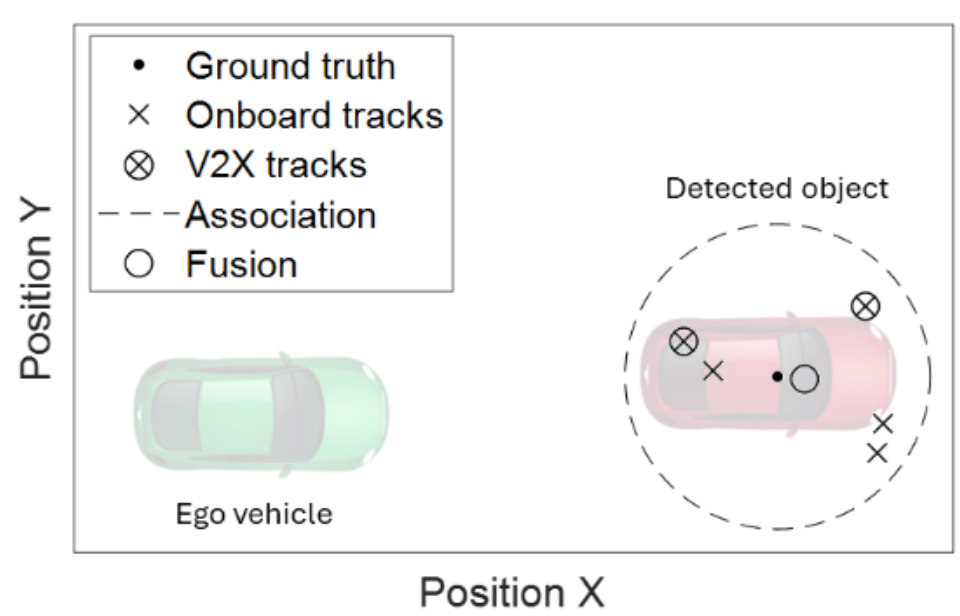

Fig. 2. Representation of ground truth, tracks (onboard and V2X), association and fusion in a 2D plane.

After association, the system performs track-to-track fusion, wherein observations of the same object are consolidated into a unified representation. This step integrates state information across sensors to generate a single, more reliable track per object. In this study, we apply a sensor-to-sensor fusion strategy, commonly referred to as memoryless fusion, meaning that fusion relies on the current estimates without considering historical track evolution. Several fusion techniques are available in the literature. One widely used approach is Weighted Average Fusion, where the fused state is computed as a weighted mean of the individual estimates, with weights reflecting each sensor's confidence level. Another method is Particle Filtering, which offers a nonparametric Bayesian approach capable of handling nonlinear dynamics and non-Gaussian noise, though at the cost of higher computational load. Finally, Covariance Intersection (CI), which provides a conservative and consistent estimate by combining state vectors and covariance matrices without requiring assumptions about inter-sensor correlations—making it particularly useful in systems where such correlations are unknown or variable. In our implementation, we use the Covariance Intersection method due to its balance between robustness and computational practicality. This method produces a fused estimate that remains consistent even under uncertain correlation conditions. Ultimately, the number of fused tracks generated by the system equals the number of clusters identified during the association phase, and ideally reflects the true number of distinct objects perceived by the vehicle's extended sensing architecture.

## III. Evaluation Methodology

We evaluate the fusion of onboard sensor data with V2X data within a 5-km highway scenario featuring four lanes (two per direction). Vehicle mobility is simulated using SUMO (Simulation of Urban Mobility), considering a traffic density of 60 vehicles per kilometer. Vehicles follow the SUMO default car-following model with varying speeds (up to 120 km/h). We analyze six V2X penetration rates ranging from 0% (no V2X data available to the ego vehicle) to 100% (all vehicles equipped with V2X). In each simulation, the ego vehicle and other CAVs transmitting cooperative perception messages are randomly selected according to the chosen penetration rate. The ego vehicle is randomly placed in the central lanes within the 1.5 km to 3.5 km segment, while CAVs may be located anywhere in the scenario. To ensure statistical robustness, we perform over 1000 independent simulations, each lasting 10 seconds. The results presented represent the mean values across all simulations.

Each vehicle is equipped with five sensors, providing a 200 m sensing range and a 360º field of view under unobstructed conditions. An object is detected if it lies within a sensor's coverage area and is not occluded by other vehicles. Each sensor employs a Kalman filter for object tracking, generating corresponding covariance matrices. Tracks are not created at the initial detection; instead, the first measurement initializes the Kalman filter for each object. Sensors generate new measurements and tracks for detected objects every 100 ms, operating synchronously. Sensors measure object position and velocity with distance-dependent noise, modeled as zero-mean Gaussian distributions applied to the ground-truth values. Two sensing noise models are considered. We consider a low sensing noise model [6] with standard deviations for the position and velocity equal to $\sigma_{low}^{pos}(d_{o,s}) = \sigma_{low}^{vel}(d_{o,s}) = d_{o,s}/1000$ , where $d_{o,s}$ is the distance between object and sensor. We also study a high sensing noise model, with the same standard deviation for the velocity error, $\sigma_{high}^{vel}(d_{o,s}) = d_{o,s}/1000$, but higher standard deviation for the position, $\sigma_{high}^{pos}(d_{o,s}) = 0.01 + d_{o,s}/20$.

All CAVs transmit cooperative perception messages every 100 ms. Each message includes the relative position, velocity, and covariance matrices of all detected objects. The periodic transmission of all detected objects aligns with the SAE J3224 standard for sensor sharing and ETSI TS 103 324 for collective perception when the *ObjectInclusionConfig* flag is *false*. Upon receiving a cooperative perception message, the ego vehicle transforms the relative positions of detected objects into absolute coordinates, using the GNSS position of the transmitting vehicle as reference. Consequently, GNSS errors introduce an additional noise source in object localization. In this work, we consider low and high GNSS errors modeled as zero-mean Gaussian distributions with standard deviations of 0.5 m and 2 m, respectively [14].

Packet losses can affect the fusion performance and are analytically modeled considering half-duplex errors, signal sensing errors, propagation losses, and packet collisions, following the model proposed in [15] forLTE-V2X PC5 mode 4. All CAVs communicate using a transmission power of 23 dBm, MCS 9 (QPSK modulation with approximately 0.7 coding rate), and utilize four sub-channels per sub-frame in the considered 10 MHz bandwidth.

## IV. Object Awareness

The performance of the fusion of onboard sensor data with V2X data is examined at varying penetration rates under low/high sensing noise and low/high GNSS errors. Increasing the penetration rate of CAVs has two significant effects on the fusion process, quantified in Fig. 3 for the considered scenario. First, increasing the number of CAVs augments the amount of information that the ego vehicle receives about each object (vehicles, in our scenario). This is illustrated in Fig. 3a, showing the average number of CAVs ($N_{CAVs}$) detecting each object. When multiple CAVs detect the same object, each transmits information about it in their cooperative perception messages. Consequently, the fusion process at the ego vehicle must handle multiple messages regarding the same object. Receiving more data about an object could help reduce uncertainty; however, noisy data may degrade the accuracy of the fusion output. Fig. 3b depicts the probability that an object is detected by at least one CAV as a function of the penetration rate. This probability depends

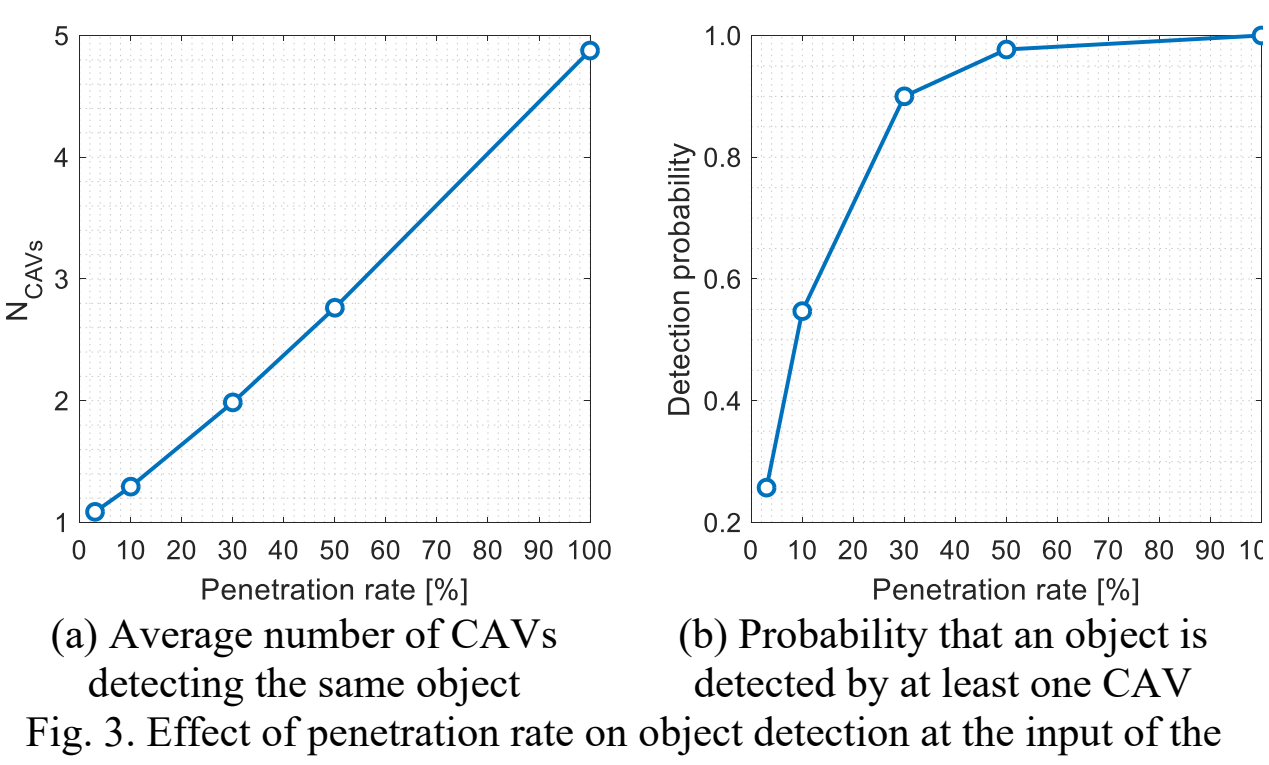

(a) Average number of CAVs detecting the same object

(b) Probability that an object is detected by at least one CAV

Fig. 3. Effect of penetration rate on object detection at the input of the fusion pipeline.

on factors such as traffic density, number of lanes, sensor detection range, and field of view. As the figure shows, the probability that an object is detected increases monotonically with saturation at 1. This probability is relevant because, if an object is not detected by any CAV, it cannot be reported in any cooperative perception message, making it unavailable to the fusion process at the ego vehicle.

To evaluate the effectiveness of fusing onboard sensor data with V2X data, we introduce the Object Awareness Rate (OAR) metric. The OAR measures the capability of the fusion process to detect existing objects (true positives). In this study, the global nearest neighbor (GNN) method is used to assign each detected object from the fusion pipeline to a corresponding ground-truth object. The GNN method assigns detections based on the minimum distance between detected and ground-truth objects, enforcing a one-to-one matching constraint. The OAR is then computed as the number of detected–ground-truth object pairs with a localization error lower than 5 meters (typical vehicle length), divided by the total number of ground-truth objects. Localization error is quantified as the Euclidean distance between the estimated position of a detected object at the output of the fusion pipeline and the ground-truth position of its corresponding assigned object. Fig. 6 illustrates the OAR as a function of the distance between the ego vehicle and detected objects, across different penetration rates, under low sensing noise and low GNSS errors. The results obtained clearly indicate that the OAR strongly depends on the penetration rate. At 0% penetration—where the ego vehicle relies solely on its onboard sensors—the OAR is relatively high at short distances but decreases rapidly as distance increases. An increase in penetration rate significantly improves the OAR, achieving over 88% detection up to 300 meters at only 30% penetration, and exceeding 97% at 50% penetration. These findings align with the results presented in Fig. 3b, demonstrating that the penetration rate is the limiting factor in scenarios with low sensing noise and GNSS errors since a considerable number of objects remain undetected by any CAV.

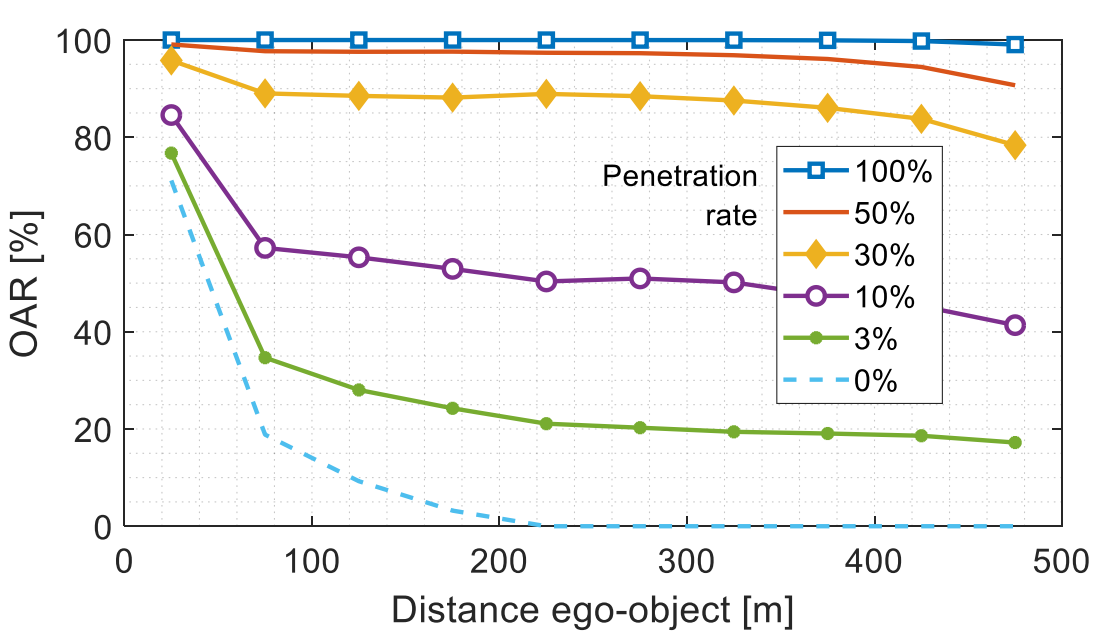

Fig. 6. OAR (Object Awareness Rate) as a function of the distance between the ego vehicle and the object for low detection noise and low GNSS errors.

Fig. 7 illustrates the spatial average of the OAR as a function of the penetration rate for the different levels of sensing noise and GNSS errors considered in this study. As shown in the figure, a higher OAR is achieved under low sensing noise and low GNSS errors, corresponding to the scenario illustrated in Fig. 6. Fig. 7 also reveals that sensing noise significantly impacts the OAR. With the high sensing noise model considered, the OAR is degraded between 10% and 24% (depending on the penetration rate) compared to the scenario with low sensing noise. The maximum OAR obtained with high sensing noise is approximately 80% at the highest penetration rate. These results demonstrate that sensing noise is a critical limiting factor affecting the effectiveness of cooperative perception and the capability of the fusion process to detect existing objects. The results in Fig. 7 further indicate that GNSS errors have only a minor impact on the OAR metric. As can be observed, scenarios with high GNSS errors reduce the OAR by only around 2% compared to those with low GNSS errors. These findings confirm that, although increasing penetration rate enhances object awareness, the effectiveness of the fusion process is highly dependent on sensing noise levels, while being less sensitive to GNSS accuracy. Overall, the results obtained indicate that sensing noise has a more substantial impact on the average OAR than GNSS errors. Low sensing noise ensures better perception performance, whereas high sensing noise introduces significant localization errors, ultimately reducing fusion effectiveness even at high penetration rates.

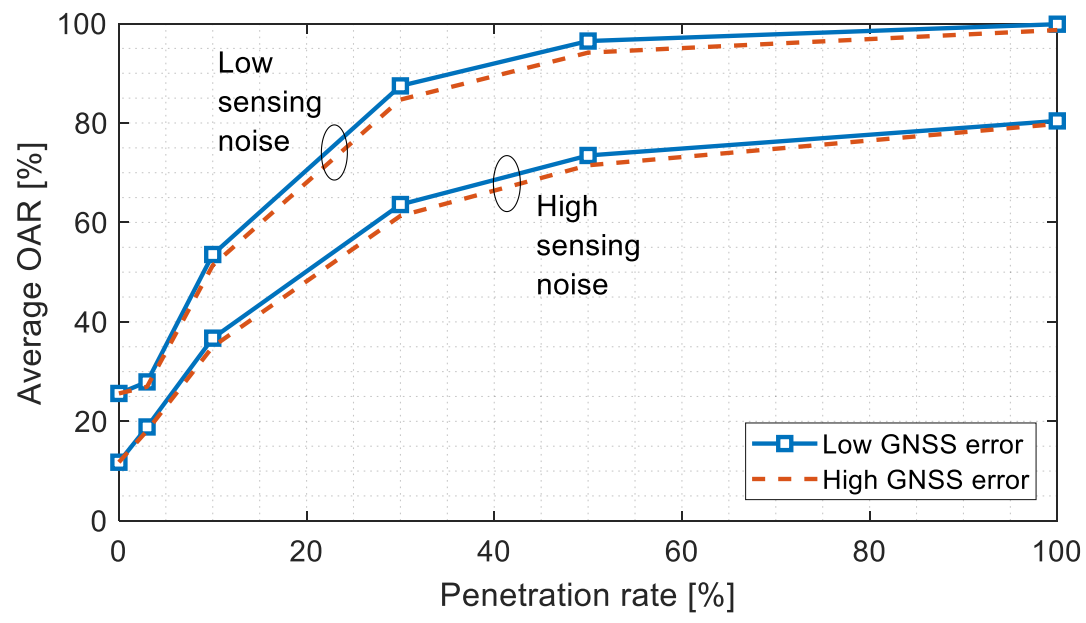

Fig. 7. Average OAR (Object Awareness Rate) as a function of the penetration rate for different levels of sensor noise and GNSS errors.

## V. Fusion Accuracy Analysis

Sensing noise and GNSS errors affect the localization accuracy at the output of the fusion pipeline of all objects, including those correctly assigned. In this study, objects are considered correctly assigned if their localization error is less than 5 meters (these objects are included in the OAR metric). Fig. 6 uses boxplots to illustrate the localization error for correctly assigned objects located within 50 meters of the ego vehicle (i.e., those most critical for immediate driving decisions). Each boxplot indicates the median and the 25th and 75th percentiles, while the whiskers represent the 5th and 95th percentiles. The figure differentiates between the various levels of sensing noise and GNSS errors considered in this paper. A common observation across all scenarios is that increasing the penetration rate leads to a higher localization error. This implies that receiving more V2X data containing erroneous information can reduce localization accuracy, even for correctly assigned objects. This illustrates a trade-off: higher penetration rates can raise OAR but may also increase localization error due to noisier V2X inputs. A detailed analysis of each scenario reveals that when sensing noise and GNSS errors are low, localization errors remain small (below 1 meter), as illustrated in Fig. 6a. Increasing the GNSS error (Fig. 6b) results in a significant rise in localization errors. This outcome highlights the influence of GNSS accuracy on localization error, even though GNSS errors have only a small effect on the OAR metric, as previously observed in Fig. 7. Finally, Fig. 6c and Fig. 6d illustrate a considerable increase in localization error when sensing noise is high. Although the median localization error remains below 1 meter (Fig. 6c) and below 2 meters (Fig. 6d), the 95th percentile exceeds 4 meters. These results emphasize that, despite an object being correctly detected (and thus included in the OAR metric), its localization accuracy remains susceptible to noise, representing an ongoing challenge for autonomous driving.

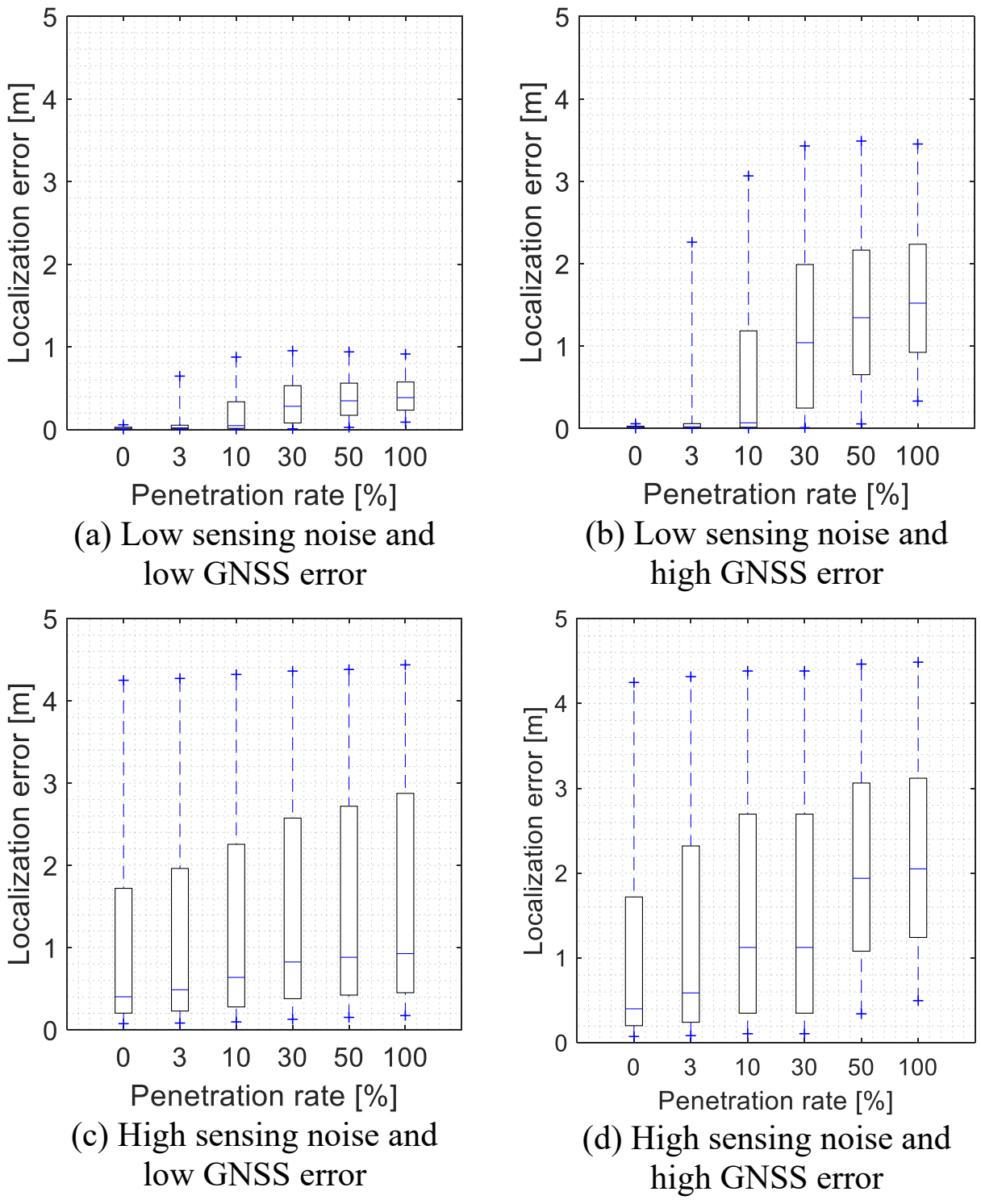

Fig. 6. Localization error for correctly assigned objects for up to 50 meters distance from ego.

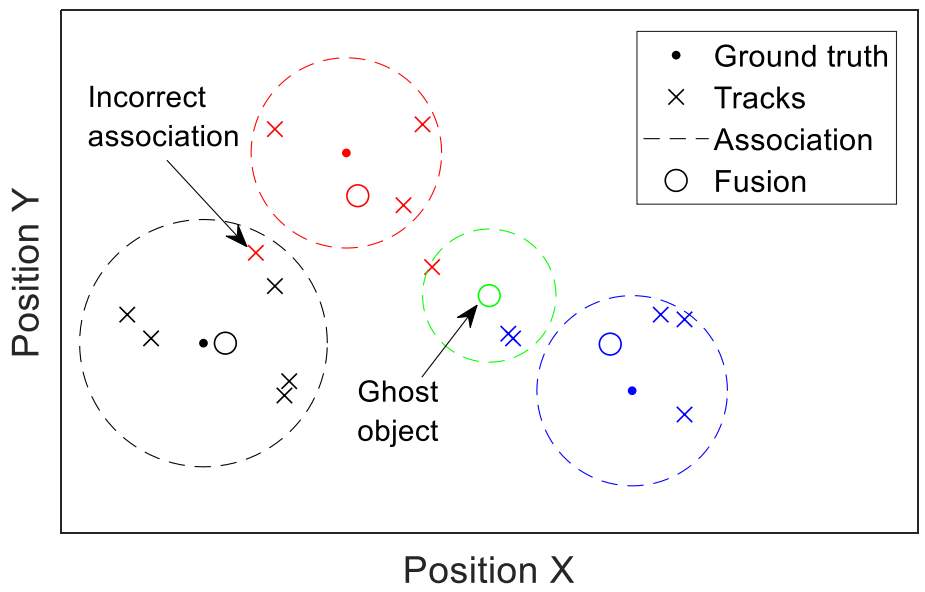

Fig. 7. Illustration of incorrect associations and ghost objects.

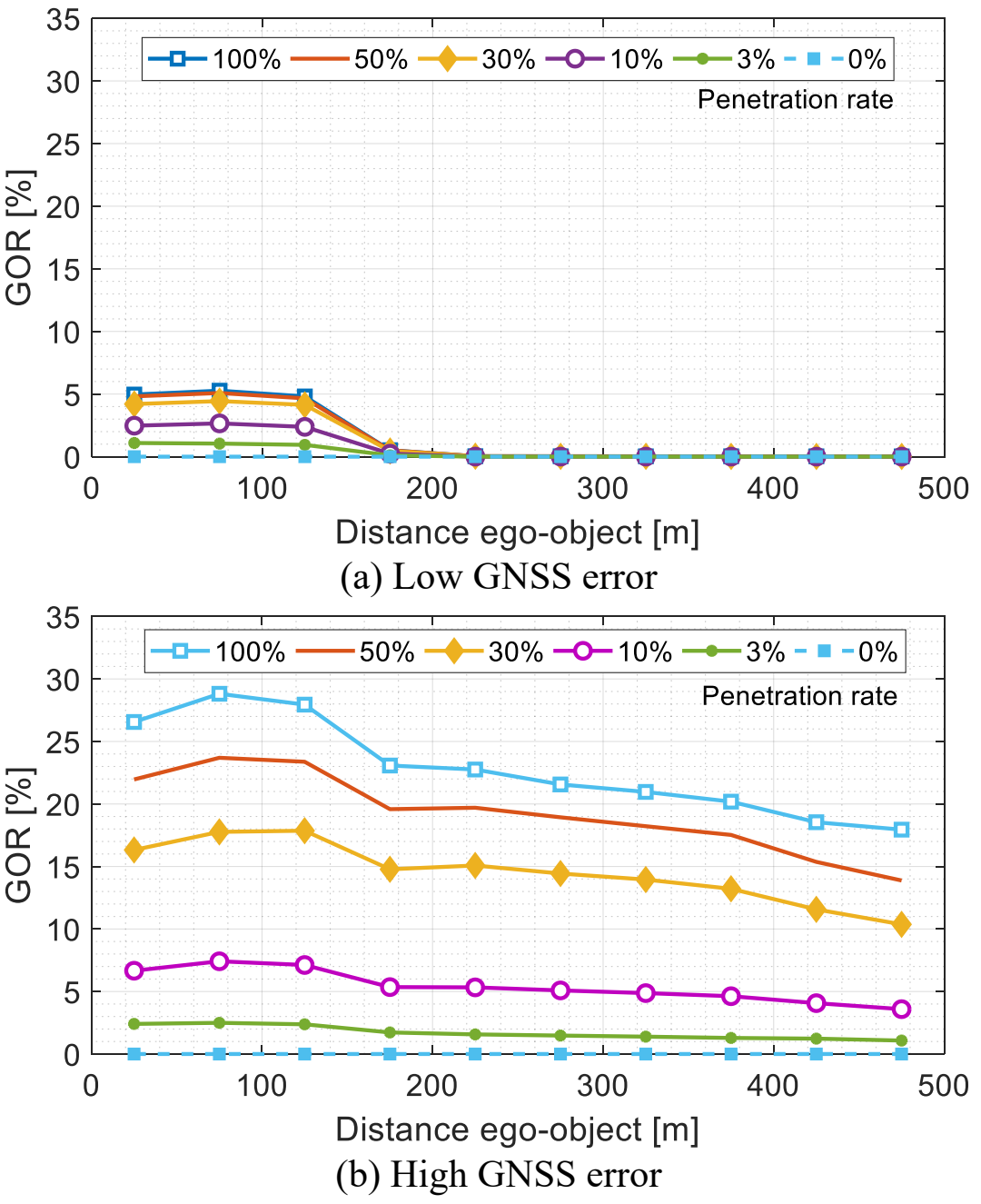

Fig. 8. GOR (Ghost Object Rate) as a function of the distance between the ego vehicle and the object for low sensing noise.

Large localization errors can reduce the capability of the fusion process to detect existing objects. This negative effect occurs when the association algorithm fails to correctly group tracks belonging to the same object prior to track-to-track fusion. In noisy environments, the association algorithm may erroneously associate tracks from different objects, as illustrated in Fig. 7, degrading the accuracy of the fused positions and thus negatively affecting the OAR, as previously demonstrated. Additionally, the association algorithm also creates ghost objects that do not really exist (false positives), which is also illustrated in Fig. 7. Here, a ghost object is any detection unassigned to ground truth or assigned with a localization error higher than 5 m. The presence of ghost objects can have an important impact on the driving decisions of the CAVs.

To quantify this negative effect, we define the Ghost Object Rate (GOR) as the number of ghost objects divided by the total number of ground-truth objects. A ghost object is defined as a detected object that either remains unassigned to any ground-truth object after the assignment process, or is assigned but exhibits a localization error greater than 5 meters. Fig. 8 illustrates the GOR as a function of the distance between the ego vehicle and the detected objects for different penetration rates under low sensing noise conditions. The figure differentiates between scenarios with low (Fig. 8a) and high (Fig. 8b) GNSS errors. In both cases, the results show that the GOR increases with the penetration rate, demonstrating that receiving more cooperative perception data may not always be beneficial if the data has inaccuracies. Fig. 8a indicates that even in the most favorable scenario (low sensing noise and low GNSS error), a non-negligible number of ghost objects is generated, particularly at short distances, which are critical for the ego vehicle. A cluster can only contain tracks from different sensors. If a received V2X track (e.g. track A) is associated with an onboard track, and is close to another V2X track (e.g. track B) from the same transmitter, they cannot be associated because they were generated by the same sensor. As a result, track B is left unmerged, and it becomes a ghost object, as illustrated in Fig. 9. The figure also shows that no ghost objects were identified beyond the sensing range of the ego vehicle, highlighting the inherent challenges associated with fusing onboard sensor data and V2X data, even under favorable conditions. Fig. 8b presents the GOR for low sensing noise combined with high GNSS errors. Comparing Fig. 8a and Fig. 8b clearly shows the significant influence of GNSS errors on the creation of ghost objects, despite GNSS errors having only a minor impact on the OAR (Fig. 7). The GOR is highest at shorter distances (within sensing range) as in Fig. 8a. Beyond the sensing range, the GOR decreases with distance due to packet losses, which reduce the amount of V2X data received from more distant objects. These results highlight the importance of mitigating false positives (ghost objects) to ensure that the fusion of onboard sensor and V2X data enhances perception without introducing errors.

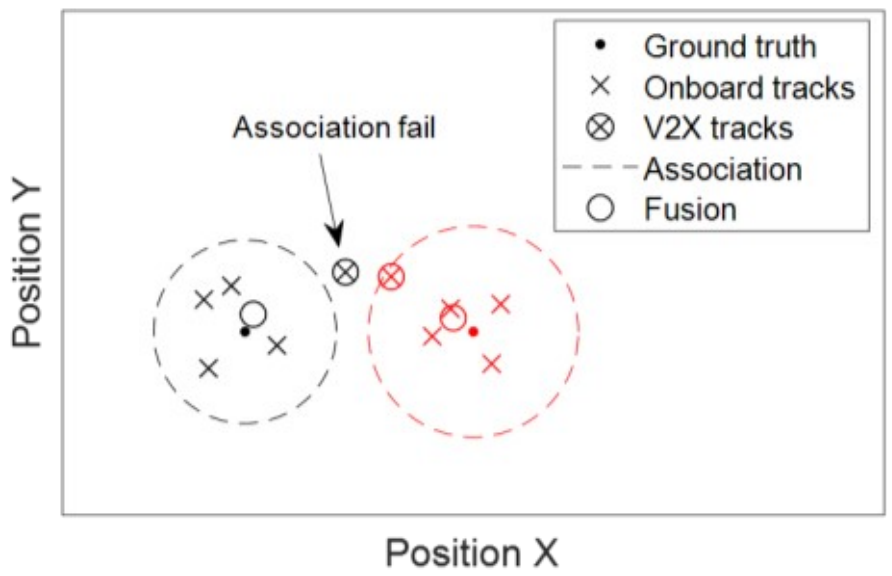

Fig. 9. Failure of association in the cluster merging logic with V2X tracks. Each colour represents a ground truth object.

Fig. 10 shows the spatial average of the GOR as a function of the penetration rate for the different levels of sensor noise and GNSS errors considered in this study. The results in Fig. 10 reveal the significant impact of GNSS errors on the GOR metric, independent of sensing noise levels. Fig. 10 also shows that with low sensing noise, increasing the penetration rate does not substantially increase the number of ghost objects. However, when sensing noise is high, an increased penetration rate can significantly raise the GOR, especially when GNSS errors are also high. These results indicate that although higher penetration rates improve the capability of the fusion process to correctly detect existing objects, they simultaneously increase the risk of generating ghost objects, particularly in scenarios characterized by significant sensing noise and GNSS errors.

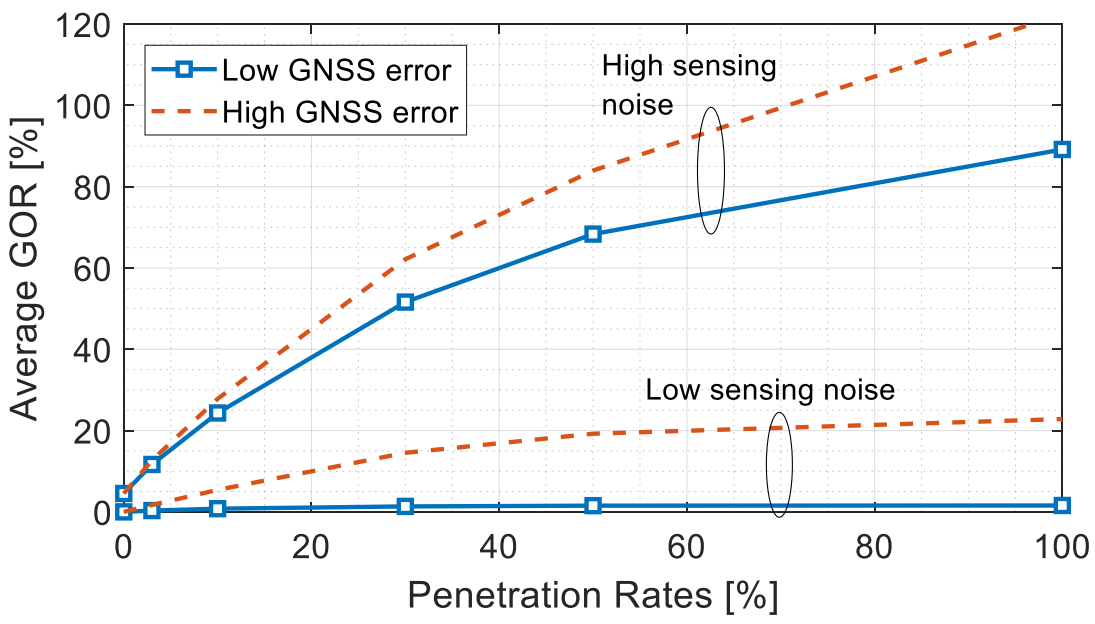

Fig. 10. Average Ghost Object Rate (GOR) as a function of the penetration rate for different levels of sensor noise and GNSS errors.

## VI. CONCLUSION

This study analyzes the fusion of onboard sensor data and V2X data in the context of cooperative perception, addressing the question of whether receiving more V2X information is beneficial or not under noisy conditions. To this end, we evaluate the fusion performance across various penetration rates and different levels of sensing noise and GNSS errors. The results show that at low V2X penetration rates, many objects are simply not detected by any CAV, and therefore no data about them is transmitted in cooperative perception messages. Increasing the penetration rate improves the capability of the fusion process to correctly detect existing objects (true positives) significantly. However, this capability may degrade in the presence of high sensing noise, which can cause the association process to fail. Our analysis also reveals that ghost objects (false positives) can emerge from the fusion process even under the most favorable conditions studied (i.e., low sensing noise and low GNSS errors). The number of ghost objects increases substantially when sensing noise and GNSS errors are high, and when the penetration rate is also high. A ghost object may trigger unnecessary braking or steering maneuvers, negatively impacting the safety, comfort, and overall performance of cooperative automated driving. These findings highlight the necessity of V2X data quality control processes, robust data association strategies and advanced fusion mechanisms to maximize the perception advantages of cooperative perception while minimizing ghost object generation.

## REFERENCES

[1] C. Pilz, *et al.*, "Collective perception: A delay evaluation with a short discussion on channel load," *IEEE Open Journal of Intelligent Transportation Systems*, vol. 4, pp. 506-526, 2023.

[2] M. Aeberhard et al., "Track-to-track fusion with asynchronous sensors using information matrix fusion for surround environment perception," *IEEE Transactions on Intelligent Transportation Systems*, vol. 13, no. 4, pp. 1717-1726, 2012.

[3] Z. Wang, Y. Wu, and Q. Niu, "Multi-Sensor Fusion in Automated Driving: A Survey," *IEEE Access*, vol. 8, pp. 2847-2868, 2019.

[4] T. Billington et al., “Enhancing track management systems with vehicle-to-vehicle enabled sensor fusion,” *Proc. IEEE 100th Veh. Technol. Conf. (VTC2024-Fall)*, pp. 1–6, Oct. 2024

[5] A. Rauch et al., “Car2x-based perception in a high-level fusion architecture for cooperative perception systems,” *Proc. IEEE Intell. Vehicles Symp.*, Alcala de Henares, Spain, pp. 270-275, Jun. 2012.

[6] M. M. De Lucena et al., "Influence of network latency on collective perception aggregation in V2X networks," *Proc. IEEE Vehicular Networking Conference (VNC)*, pp. 1-8, May 2024.

[7] A. Figueiredo et al., "Enhancing Vehicular Network Efficiency: The Impact of Object Data Inclusion in the Collective Perception Service," *IEEE Open Journal of Intelligent Transportation Systems*, vol. 5, pp. 454-468, Aug. 2024.

[8] G. Thandavarayan et al., "Scalable Cooperative Perception for Connected and Automated Driving", *Journal of Network and Computer Applications*, vol. 216, pp. 103655, May 2023

[9] A. H. Sakr, "Evaluation of redundancy mitigation rules in V2X networks for enhanced collective perception services," *IEEE Access*, vol. 12, pp. 137696 – 137711, Sept. 2024.

[10] M. Aeberhard and N. Kaempchen, "High-level sensor data fusion architecture for vehicle surround environment perception," in *Proc. 8th Int. Workshop Intell. Transp.*, vol. 665, Mar. 2011, pp. 1-7.

[11] S. Julier and J. K. Uhlmann, "General decentralized data fusion with covariance intersection," in *Handbook of Multisensor Data Fusion*, CRC Press, pp. 339-364, 2017.

[12] A. Houenou, P. Bonnifait, V. Cherfaoui, and J. F. Boissou, "A track-to-track association method for automotive perception systems," in *Proc. IEEE Intelligent Vehicles Symposium*, pp. 704-710, June 2012.

[13] L. M. Wolf, et al., "Track-to-track association based on stochastic optimization," *Proc. 26th International Conference on Information Fusion (FUSION)*, Charleston, SC, USA, 27-30 June 2023.

[14] M. M. S. Alghananim and W. Y. Ochieng, “Maximum non-bounded difference method for overbounding Global Navigation Satellite System errors,” *GPS Solutions*, vol. 29, no. 1, pp. 1–14, 2025.

[15] M. Gonzalez-Martín, et al., "Analytical Models of the Performance of C-V2X Mode 4 Vehicular Communications," *IEEE Transactions on Vehicular Technology*, vol. 68, no. 2, pp. 1155-1166, Feb. 2019.